\documentclass[conference]{IEEEtran}
\IEEEoverridecommandlockouts
\usepackage{cite}
\usepackage{amsmath,amssymb,amsfonts}
\usepackage{graphicx}
\usepackage{textcomp}
\usepackage{xcolor}
\usepackage{makecell}
\def\BibTeX{{\rm B\kern-.05em{\sc i\kern-.025em b}\kern-.08em
    T\kern-.1667em\lower.7ex\hbox{E}\kern-.125emX}}

\graphicspath{ {./figures/} }
\usepackage{algorithm}		
\usepackage[noend]{algpseudocode} 
\usepackage{url}            
\begin{document}

\title{AIPerf: Automated machine learning as an AI-HPC benchmark\\
}

\author{\IEEEauthorblockN{1\textsuperscript{*} Zhixiang Ren}
\IEEEauthorblockA{\textit{} \\
\textit{Peng Cheng Laboratory}\\
renzhx@pcl.ac.cn}
\and
\IEEEauthorblockN{2\textsuperscript{nd} Yongheng Liu}
\IEEEauthorblockA{\textit{} \\
\textit{Peng Cheng Laboratory}\\
yongheng.liu@pcl.ac.cn}
\and
\IEEEauthorblockN{3\textsuperscript{rd} Tianhui Shi}
\IEEEauthorblockA{\textit{Department of Computer Science and Technology} \\
\textit{Tsinghua University}\\
sth19@mails.tsinghua.edu.cn}
\and
\IEEEauthorblockN{4\textsuperscript{th} Lei Xie}
\IEEEauthorblockA{\textit{Department of Computer Science and Technology} \\
\textit{Tsinghua University}\\
xie-l18@mails.tsinghua.edu.cn}
\and
\IEEEauthorblockN{5\textsuperscript{th} Yue Zhou}
\IEEEauthorblockA{\textit{} \\
\textit{Peng Cheng Laboratory}\\
zhouy@pcl.ac.cn}
\and
\IEEEauthorblockN{6\textsuperscript{th} Jidong Zhai}
\IEEEauthorblockA{\textit{Department of Computer Science and Technology} \\
\textit{Tsinghua University}\\
zhaijidong@tsinghua.edu.cn}
\and
\IEEEauthorblockN{7\textsuperscript{th} Youhui Zhang}
\IEEEauthorblockA{\textit{Department of Computer Science and Technology} \\
\textit{Tsinghua University}\\
zyh02@tsinghua.edu.cn}
\and
\IEEEauthorblockN{8\textsuperscript{th} Yunquan Zhang}
\IEEEauthorblockA{\textit{Institute of Computing Technology} \\
\textit{Chinese Academy of Sciences}\\
zyq@ict.ac.cn}
\and
\IEEEauthorblockN{\textsuperscript{*} Wenguang Chen}
\IEEEauthorblockA{\textit{Department of Computer Science and Technology} \\
\textit{Tsinghua University}\\
cwg@tsinghua.edu.cn}
}

\maketitle

\begin{abstract}
The plethora of complex artificial intelligence (AI) algorithms and available high performance computing (HPC) power stimulates the expeditious development of AI components with heterogeneous designs.
Consequently, the need for cross-stack performance benchmarking of AI-HPC systems emerges rapidly.
The de facto HPC benchmark LINPACK can not reflect AI computing power and I/O performance without representative workload.
The current popular AI benchmarks like MLPerf have fixed problem size therefore limited scalability.
To address these issues, we propose an end-to-end benchmark suite utilizing automated machine learning (AutoML), which not only represents real AI scenarios, but also is auto-adaptively scalable to various scales of machines.
We implement the algorithms in a highly parallel and flexible way to ensure the efficiency and optimization potential on diverse systems with customizable configurations.
We utilize operations per second (OPS), which is measured in an analytical and systematic approach, as the major metric to quantify the AI performance.
We perform evaluations on various systems to ensure the benchmark's stability and scalability, from 4 nodes with 32 NVIDIA Tesla T4 (56.1 Tera-OPS measured), up to 512 nodes with 4096 Huawei Ascend 910 (194.53 Peta-OPS measured), and the results show near-linear weak scalability.
With flexible workload and single metric, our benchmark can scale and rank AI-HPC easily.

\end{abstract}


\section{Introduction}
\label{sec:intro}
\begin{figure}[htbp]
	\centering
	\includegraphics[width=0.3\textwidth]{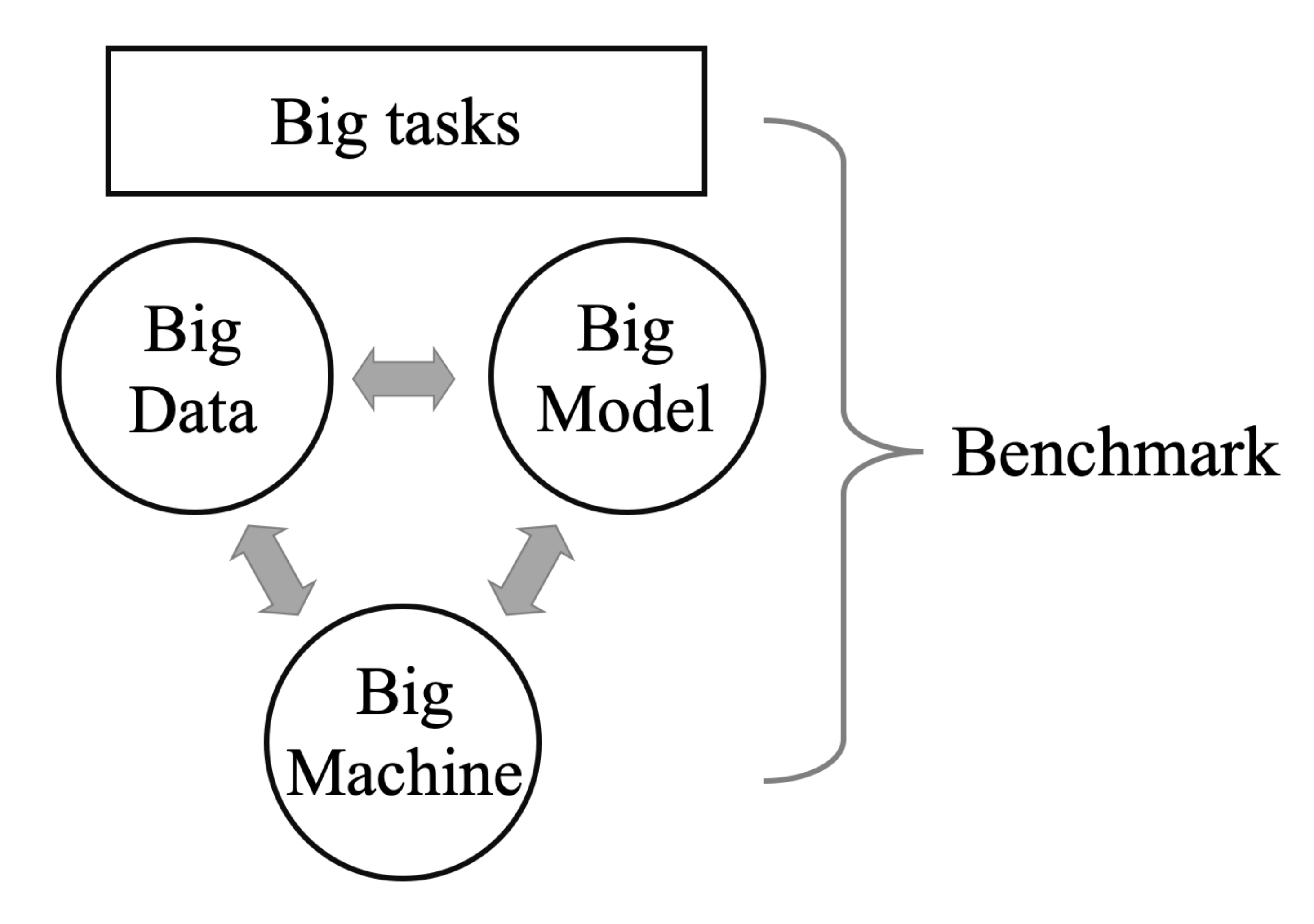}
	\caption{The converge of AI and HPC with the growth of model, machine, data, as well as potential tasks. The benchmark should cover the system heterogeneity and reflect the cross-stack performance.}
	\label{fig:ai-hpc-overview}
\end{figure}

Artificial intelligence (AI), machine learning (ML) and deep learning (DL) have drawn tremendous attention in recent years. 
DL requires a training process~\cite{dl}, which is essentially a multi-dimensional fitting, to automatically adjust the weights (parameters) of the neural network. 
As the learnable data grows at an unprecedented rate, the high performance computing (HPC) machines are needed for the large AI model to harness the big data and extract the complex abstractions~\cite{ai-hpc-convergence-1}.
The hybrid HPC models with AI surrogates reveal a collection of unique and novel opportunities for scientific breakthrough and unforeseeable discoveries~\cite{ai-hpc-convergence-3-scientific}, as well as business innovations and other societal benefits.
The increase in algorithmic advances of AI algorithms, available computing power and data collections, as well as the demand for scalable and data-driven solutions stimulate the convergence of AI and HPC machines~\cite{ai-hpc-convergence-2}. 
The convergence~\cite{summit-ai} still faces multiple challenges, like the effective and parallel implementation of algorithms on large scale clusters, high bandwidth as well as low latency communications between distributed workers, and high-speed interconnections to the network file system. 
The HPC systems need to incorporate the support for various AI workloads on top of inconsistent accelerators and software frameworks for an AI-HPC adaption.
Consequently, the need for an open and reliable benchmark suite to comprehensive evaluate the cross-stack performance of heterogeneous AI-HPC systems emerges rapidly, as shown in Fig.~\ref{fig:ai-hpc-overview}.

There are three major challenges for AI-HPC benchmarking.
First, the benchmark workload needs to represent the real problems running on AI-HPC, regarding hardware utilization, set-up cost and computing patterns etc.
Second, the benchmark workload is preferably auto-adaptive to various scales of machines without extra human effort.
Third, define a simple system metric on AI performance and design an approach to measure it.
Unfortunately, the current HPC and AI benchmarks do not address these challenges.
The de facto HPC benchmark LINPACK does not measure the cross-stack performance on AI without a representative workload.
On the other hand, popular AI benchmarks like MLPerf~\cite{ai-benchmark-mlperf} do have representative AI workloads, but they have a fixed workload size and are often used to benchmark small systems.
Moreover, they does not scale automatically with machines and require a lot human effort for tunning.
This is fallacious since the increased computing power tends to be utilized to attack larger problems instead of the same problem with less time.
The fixed problems can also not adapt to different scales of machines automatically.

Automated machine learning (AutoML) can search and optimize the AI models more automatically and is getting increasing attention in the AI community.
As an representative AI application, AutoML contains basically all the critical components regarding the primary computing operations (e.g. sparse matrices multiplication), calculation precision (in FP-32 or lower) and workflow in real AI scenarios.
More importantly, AutoML can be implemented in a flexible style so that it scales automatically with the number of machines.
Besides, the pseudo-random generated architecture and extreme computational cost would address the evolving and diverse architectural designs in AI research and fully push the system limits in benchmarking.
Considering all the advantages, AutoML is a desired workload and we choose it to tackle the first two challenges.
For the third challenge, we learn from the success of LINPACK and Top500 and utilize operations per second (OPS) as our benchmark score to quantitatively measure the AI computing power.
The OPS is measured in an analytical and systematic method to account for both training and inference processes.
With the auto-adaptive workload and single metric measurement, our benchmark can easily rank the various size of machines from small clusters to large AI-HPC.

In summary, our main contributions are the following:
\begin{itemize}
	\item We propose AutoML as a representative and auto-adaptive workload to establish an end-to-end benchmark suite for AI-HPC.
	\item Our implementation is highly parallel and customizable to keep the optimization potential on diverse systems.
	\item We propose an analytical and speedy approach to calculate the operation rate of neural networks with different architectures.
	\item We evaluate our benchmark on various systems to ensure the benchmark's scalability and stability.
\end{itemize}

The rest of this paper is organized as follows. 
In Section~\ref{sec:rel}, we review the existing HPC and AI benchmarks and point out their downsides for AI-HPC benchmarking. 
In Section~\ref{sec:bkg}, we briefly review AutoML and the popular frameworks for AI. 
In Section~\ref{sec:meth}, we describe the details of our algorithms, implementations and measurements.
In Section~\ref{sec:eva}, we evaluate our benchmark on different scales of machines. 
We summarize our work in Section~\ref{sec:con}.
The source code, specifications and detailed procedures are publicly accessible on GitHub\footnote{AIPerf: https://github.com/AI-HPC-Research-Team/AIPerf}.

\section{Related Work}
\label{sec:rel}

\subsection{HPC Benchmarks}
\label{sec:rel_hpc}
LINPACK~\footnote{https://www.top500.org/project/linpack/} is the popular HPC benchmark nowadays.
It is essentially an algebra library that solves a dense system of linear equations that is the heart of many computational science problems.  
There are three reasons why LINPACK is not suitable for benchmarking AI-HPC.
First, the problem size is usually manually decided and can not be scaled automatically based on the tested machines. 
Second, LINPACK provides little information about the set-up cost and I/O ability, which are critical data-intensive applications like AI.
This is problematic since most algorithms do more data motion than arithmetic~\cite{hpc-benchmark-hint}.
Third, the calculation is performed in FP-64, while most AI applications typically only require FP-32 or even FP-16.
HPL-AI Mixed-Precision benchmark~\cite{hpc-benchmark-hplai} is developed based on LINPACK to highlight the third issue, but it still suffers from the other two issues.
Other HPC benchmarks including NASA Parallel Benchmarks~\footnote{https://www.nas.nasa.gov/publications/npb.html}, SLALOM~\cite{hpc-benchmark-slalom} and HINT~\cite{hpc-benchmark-hint} do not utilize workloads that can represent real AI scenarios, therefore share the same problems as LINPACK.
Though we can not use the existing HPC benchmarks for AI-HPC, they still inspire us in the benchmark design.
For example, the biggest challenge in benchmarking is to create a single workload that can capture all the features of real applications and be auto-adaptive without a fixed problem size. 
Also, further performance optimization with customizable configuration is encouraged, as long as the user does not specialize the program to input data.
Last but not least, a single number metric is preferred for easy comparison and ranking.

\subsection{AI Benchmarks}
\label{sec:rel_ai}
Fair and inclusive comparison of machine computing power on AI applications is not trivial. 
As the opposite of monoculture, the system's heterogeneity, and the variety of AI workloads as well as the stochastic nature of approaches makes the benchmarking complicated.
Previous AI benchmarks attempt to highlight the challenges by incorporating different hardware systems~\cite{ai-benchmark-dnnmark,ai-benchmark-benchip,ai-benchmark-aibenchmark,ai-benchmark-acceleratorbenchmark,ai-benchmark-deepbench}, software frameworks~\cite{ai-benchmark-dlbs} or AI algorithms~\cite{ai-benchmark-fathom,ai-benchmark-aibench,ai-benchmark-aimatrix,ai-benchmark-aiia}.
More recently, end-to-end benchmarks include~\cite{ai-benchmark-dawnbench,ai-benchmark-tbd,ai-benchmark-deep500,ai-benchmark-mlperf,ai-hpc-benchmark-hpcai500} are developed to evaluate hardware systems and AI algorithms simultaneously. 
MLPerf~\cite{ai-benchmark-mlperf}, the arguably most accepted AI benchmark so far, uses time-to-accuracy to measure the co-performance of hardware and software.
This metric is an indirect quantification of the computing ability comparing to OPS, which is our metric.
Since MLPerf is composed of multiple micro-tasks, each one would result in a different measurement.
Though this approach makes the benchmark more accurate on various applications, it also makes the comparison and ranking more difficult.
Also, the limited workloads in MLPerf have insufficient scalability with fixed problem size.
Other AI benchmarks have similar drawbacks as MLPerf.
Overall, there are two reasons why the existing AI benchmarks are not suitable to be AI-HPC benchmark: 

\begin{itemize}
	\item Existing AI benchmarks have fixed problem size therefore limited scalability.
	\item Existing AI benchmarks do not provide a single and direct measurement to quantify performance.
\end{itemize}

\section{Background}
\label{sec:bkg}

\subsection{Automated Machine Learning}
\label{sec:bkg_automl}
\begin{figure}[htbp]
	\centering
	\includegraphics[width=0.4\textwidth]{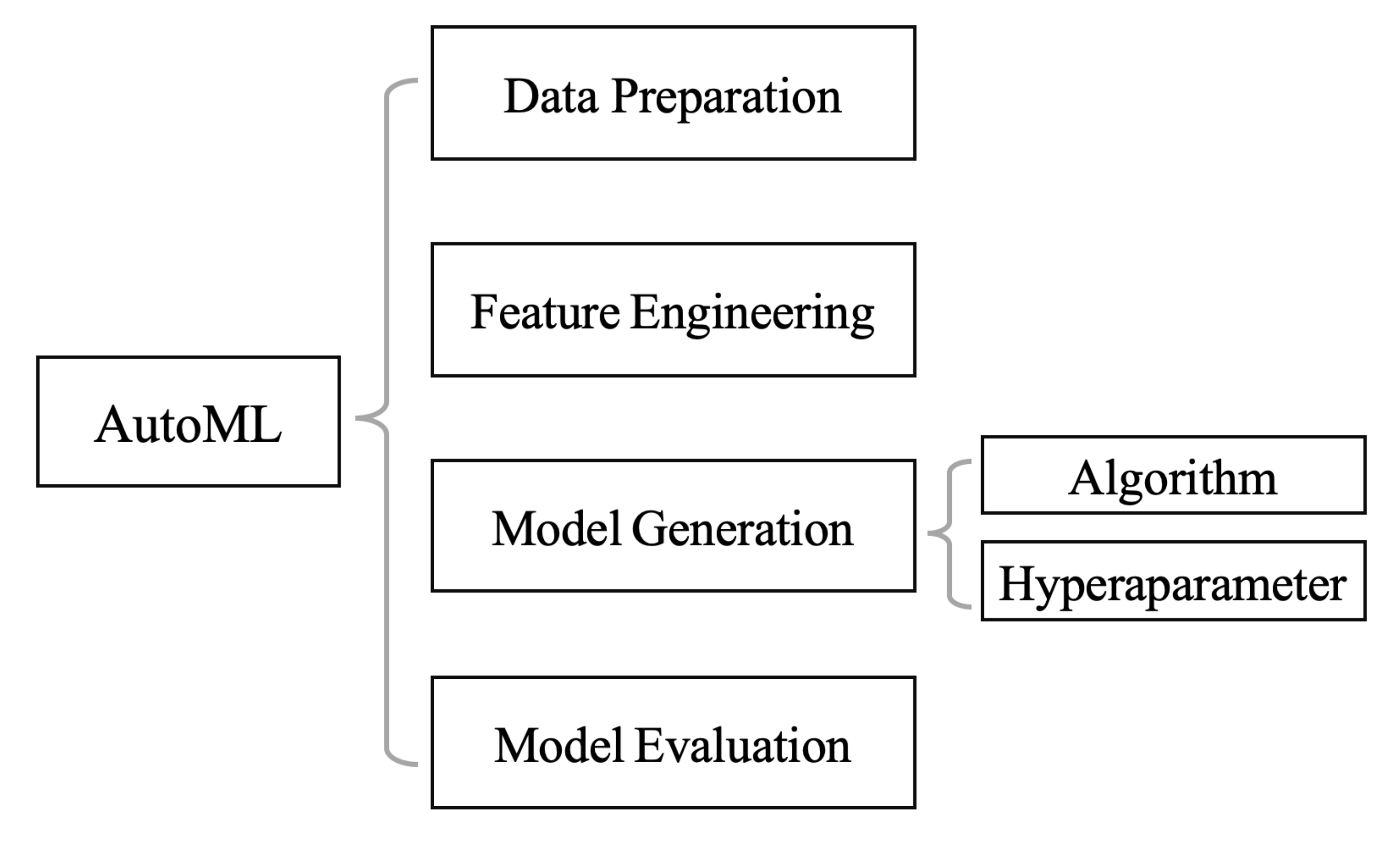}
	\caption{An overview of the AutoML. We limit our attention to the model generation in this paper.}
	\label{fig:automl}
\end{figure}

Developing AI solutions have mostly relied on a complex model design which involves human expertise heavily and is extremely time-consuming.
To explore the architecture space more efficiently and optimize the model automatically, AutoML~\cite{automl-book} emerges as the AI model complexity increases exponentially~\footnote{OpenAI: https://openai.com/blog/ai-and-compute/} in recent years.
It may sound surprising but AutoML is already mature enough to rival human experts to make a real impact on AI research.
Overall, AutoML is inherently computing-intensive, highly scalable and representative of AI-like workflows. 
Considering all the unique advantages, we choose AutoML as our benchmark workload.
As shown in Fig.~\ref{fig:automl}, AutoML contains various parts~\cite{automl-review}. 
The first part is data preparation, which involves data collection and data cleaning. 
The second part is feature engineering, including feature selection, feature construction and feature extraction. 
Although data and features lay the foundations of AI performance, they depend on the application scenarios and are irrelevant to the machine computing power, therefore not considered in our benchmark.
The third part is to generate the neural architecture and the optimal configuration (referred to as hyperparameters), which can have a significant impact on the performance.
The two main approaches for model generation are the experts' manual design and the automated neural architecture search (NAS~\cite{automl-nas-review}). 
Without human intervention, NAS has the potential to generate novel architectures beyond imagination and can boost the performance significantly.
Hyperparameter optimization (HPO~\cite{automl-hpo-rs}) is essentially the optimization of the loss function over the complex configuration space.
The NAS and HPO can be implemented in a parallel manner to fully utilize the distributed resources. 
Finally, model evaluation measures the performance once the candidate model is generated. 
The simplest method is to conduct the inference on the test dataset for enough epochs.
This is prohibitively expensive since there are numerous configurations for each neural architecture.
In this paper, We use warm-up and early stopping strategies~\cite{sk-learning-early-stop} that stops the training once the validation loss flats and this can provide measurements quickly to a certain degree of accuracy.

\subsection{Frameworks}
\label{sec:bkg_fm}

\paragraph{Deep Learning Frameworks}
DL frameworks provide user-friendly API and transform programs in high-level languages into an internal representation of certain functionalities.
The low-level efficient libraries, e.g. cuDNN, are invoked to execute primary operations like matrix multiplication.
Multiple solutions with desired performance exist~\cite{DL-book}, therefore implementation and customized setups vary while maintaining similar results. 
The difference is critical as the training process is stochastic and approximate intrinsically.
An open-source framework with enough community support would be a decent candidate for building the benchmark.
According to GitHub~\footnote{Deep learning frameworks: https://github.com/topics/deep-learning}, the most popular deep learning frameworks are TensorFlow~\cite{framework-tf}, Keras~\cite{framework-keras} and PyTorch~\cite{framework-pytorch}. 
TensorFlow is an open-source library for low-level numerical calculation with static computational graphs where operations are written as high-performance C++ binaries with high-level Python abstractions.
Keras is a high-level library wrapper that is built on top of frameworks like TensorFlow and provides off-the-shelf but often inflexible models.
PyTorch utilizes dynamic computation graphs that are modifiable at run-time but the "Pythonic" nature makes it less efficient for benchmarking purposes.
After carefully comparing different frameworks~\cite{framework-dl-1,framework-dl-2}, we choose TensorFlow in our benchmark evaluation for the following reasons:

\begin{itemize}
	\item TensorFlow is the most popular open-source deep learning framework so far, with a large and active community supported by Google for quick updates and frequent releases.
	\item TensorFlow is efficient, user-friendly and easy-to-debug (with TensorBoard) regarding the numerical computations for both research and deployment. 
	\item TensorFlow supports various systems with high performance and scalability.
\end{itemize}

\paragraph{AutoML Frameworks}
Various work has been done to develop user-friendly AutoML frameworks~\cite{framework-automl-1,framework-automl-2} including Neural Network Intelligence (NNI), Tree-based Pipeline Optimization Tool (TPOT) and auto-sklearn. 
NNI~\footnote{https://www.microsoft.com/en-us/research/project/neural-network-intelligence/} is a popular open-source toolkit that automates the DL model design process.
One key feature is the rich collection of algorithms to generate neural architectures and optimizing hyperparameters, as well as a simple interface for more user-defined algorithms.
Other frameworks focus on the AutoML pipeline optimization, especially data pre-processing and feature engineering, which is irrelevant for benchmarking the computing power.
Therefore we choose to build our own benchmark suite on top of NNI.

\section{Methodology}
\label{sec:meth}

\subsection{Neural Architecture Search}
\label{sec:meth_nas}
Notable successes of neural architecture designs \cite{nas-vgg,nas-googlenet,nas-resnet,nas-denselynn,nas-mobilenet,nas-senet} in the past few years have drawn enormous attention in AI research community. 
The manual design of neural architecture requires tremendous human effort, sometimes even domain knowledge in an ad-hoc fashion.
In contrast, the architectures are automatically generated by selecting and combining primary operations (e.g. convolution) with NAS approaches which can be categorized into three abstraction levels~\cite{automl-nas-review}: search space, search strategy and performance estimation strategy.
The major search strategies (algorithms)~\cite{automl-review} include random search~\cite{automl-nas-rs}, reinforcement learning~\cite{automl-nas-rl}, evolutionary~\cite{automl-nas-evol}, Bayesian optimization~\cite{automl-nas-bo} and gradient-based method~\cite{automl-nas-gb}.
Research around NAS is typically exploring three dimensions of abstractions simultaneously using various algorithms to search for different combinations of building blocks.
In the spirit of transfer learning and knowledge inheritance, \cite{automl-nas-morphism-1} proposed network transformation that transforms a pre-trained parent network to a more complex child network while preserving the input and output consistency.
The knowledge represented by the neural architecture is transformed from the parent network to the child network. 
\cite{automl-nas-morphism-2} first dubbed "network morphism" that can perform multiple transformation operations including width, depth, kernel size and skip operation.
\cite{automl-nas-morphism-3} proposed an open-source framework (Auto-Keras), which is part of NNI, to perform network morphing guided by Bayesian optimization.
Though every method has its own advantages, we choose the implementation in \cite{automl-nas-morphism-2} as our baseline for developing the benchmark.
We choose residual network~\cite{nas-resnet} (ResNet-50) as the initial model since ResNet-50 is one of the de facto showcase models in the current DL community and contains basically all the AI-related computation primitives.
We modify the morphism so that each transformation step adds a block (convolutional layer, batch normalization~\cite{batch-norm} and activation function all together) instead of just one layer.
In addition, We adapt this implementation to suit benchmarking in a parallel and distributed way which is explained later.

\subsection{Hyperparameter Optimization}
\label{sec:meth_hpo}
HPO problems can be viewed as the identification of optimal model configurations of all related hyperparameters.
Similar to NAS, HPO has three abstractions~\cite{automl-book}: search space, search approach and evaluation method.
Various search approaches can be applied to select the best hyperparameter combinations including grid search~\cite{automl-hpo-grid}, random search~\cite{automl-hpo-rs}, Bayesian optimization~\cite{automl-hpo-bo-tpe} and heuristic search like evolutionary~\cite{automl-hpo-evol} etc.
In our case, the search space is defined by the hyperparameters that are more directly related to the computational cost including the batch size and kernel size to reduce the randomness for benchmarking purposes.
We use the stochastic gradient descent (SGD) with momentum~\cite{config-optimizer} as the optimizer since it requires less memory and be more efficient.
We evaluated the different optimization approaches and then compare the validation accuracy on the test dataset. 
The results of multiple experiments on CIFAR10 show that Bayesian optimization (TPE) slightly outperforms other methods.
Similar to NAS, we use this fixed algorithm to optimize the batch size and kernel size simultaneously.
In our benchmark workflow, the HPO is performed separately after the NAS process on each worker.

\subsection{Workflow}
\label{sec:meth_wf}
\begin{figure}[htbp]
	\centering
	\includegraphics[width=0.5\textwidth]{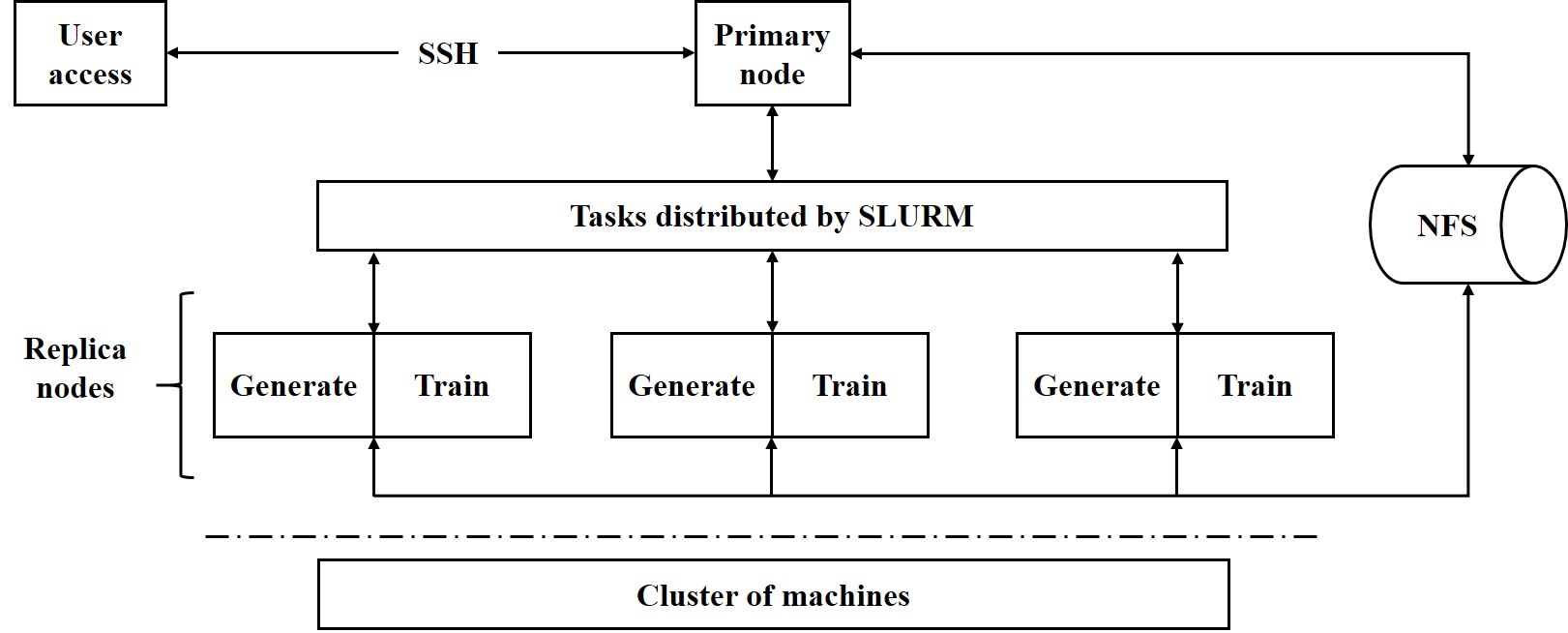}
	\caption{Schematic diagram of the benchmark workflow. The details are explained in the main text.}
	\label{fig:framework-overview}
\end{figure}

As mentioned, we choose NNI (V1.5) as a baseline to adapt to our benchmark suite.
The original NNI framework is implemented with a "primary-replica" architecture and performs the NAS and HPO on the primary node, which is the bottleneck on large clusters.
Also, not all operations in AutoML run on AI accelerators, like model generation and data movement. 
Consequently, the AI accelerator idles because of the potential bottleneck on CPU or disk I/O.
In addition, the model generation is time-consuming and can be implemented with thread parallelism on CPUs.
To address these problems and fully appreciate all computing resources in a balanced way, we need to effectively distribute the computations and use proper parallelism~\cite{dml-survey} on both CPU and AI accelerator. 
Therefore, we modify the NNI framework in various aspects as shown in Fig.~\ref{fig:framework-overview}, including performing the model generation and training on replica nodes asynchronously, utilize replica nodes' CPUs parallelly to generate new architectures and perform training parallelly with all available AI accelerators on each replica node.
We utilize data parallelism with synchronous all-reduce strategy so that all AI accelerators can train on different partitions of data and results in individual gradients, which are then aggregated all-together at each step.
We summarize our benchmark workflow as follows: 
\begin{itemize}
	\item User accesses the primary node through Secure Shell (SSH), collects information about replica nodes and creates a SLURM configuration script.
	\item The primary node dispatches workloads with SLURM to replica nodes corresponding to the requested and available resources, parallelly and asynchronously.
	\item The replica nodes receive the workloads and perform architecture searching and model training parallelly. 
	\item The CPUs on replica nodes search for new architectures based on the current historical model list, which contains detailed model information and accuracy on the test dataset, then store the candidate architecture in the buffer (e.g. network file system) for later training.
	\item The AI accelerators on replica nodes load the candidate architecture and data, utilize data parallelism to train along with HPO and then store the results in the historical model list.
	\item The running terminates once the condition is satisfied (e.g. reaching user-defined time). The final results are calculated based on the recorded metrics and then reported.
\end{itemize}

\subsection{Measurement}
\label{sec:meth_mea}
Floating-point operations per second (FLOPS) or OPS is the most cited performance metric to reflect the overall computing ability of HPC~\footnote{Top500 Project: https://www.top500.org/} as yet.
Our benchmark utilizes OPS as the major metric (score) to directly describe the computing power of AI accelerators.
Since the processing time can be easily recorded, we only need to count the total operations, and all computations are required to be conducted with floating points of at least FP-16 precision.
Toolkit like NVIDIA profiling tools (nvprof~\footnote{CUDA Toolkit Documentation: https://docs.nvidia.com/cuda/profiler-users-guide/index.html}) can record the executed operation count by kernel replay, which is exceptionally slow.
This method is also limited to NVIDIA hardware and is not suitable for various platforms.
Inspired by LINPACK, we treat the operation counting as a mathematical problem to calculate the operation needed to finish the complex computation in the training and validation processes without any optimization.
For a given dataset and model with specific hyperparameters, the theoretical operation needed to finish the training or validation is predetermined. 
If the hardware or software has any special optimization, the operation count is reduced or the execution is faster, therefore higher OPS eventually.

To calculate the operation count analytically, we need to understand the training and validation process.
DL libraries like TensorFlow use computational graphs to represent the computations and guide the workflow.
A computational graph is a directed acyclic graph where nodes represent variables or operations and edges represent function arguments (data dependency). 
Each computation is essentially a node so that variables feed values into operations and operations feed the outputs into other operations.
Computational graphs can compose complex models with simple functions and enable automatic differentiation to train the neural networks.
Backpropagation~\cite{backprop} is a reverse mode automatic differentiation~\cite{auto-diff} which applies the chain rule efficiently and recursively to compute gradients of inputs and parameters and other intermediates along with computational graphs.
As shown in Algorithm~\ref{alg:bp}, backpropagation has two parts: forward pass (FP) that compute results of operations and save intermediate values needed for gradients computation in memory and backward pass (BP) that apply the chain rule to compute the gradients of the loss function with respect to the inputs (multiply Jacobian matrices by gradients).

\begin{algorithm}
	\caption{Backpropagation \cite{backprop}}\label{alg:bp}
	\begin{algorithmic}
		\State{Forward Pass:}
		\State{1. Define the computational graph where each node represent a variable (parameters and intermediates).}
		\State{2. Visit each node in topological order to compute the variables with corresponding operations and store the values at the nodes.}
		\State{Backward Pass:}
		\State{3. Initialize the loss gradients $dL \over dy$ and all local partial derivatives$dy \over dx_i$.}
		\State{4. Visit each node in reverse topological order to compute the loss gradients w.r.t. local variables with chain rule: ${dL \over dx_i} = {dL \over dy} \times {dy \over dx_i}$.}
		\State{Return: $dL \over dx_i$ for all variables.}
	\end{algorithmic}
\end{algorithm}

The total operation count is the sum of that in FP and BP, which includes operations to calculate the gradients and the operations to update the parameters with gradient descent.
Most computations in neural networks are matrices multiplication, which is dot products $y = w_0 \times x_0 +  w_1 \times x_1 + \dots + w_{n-1} \times x_{n-1}$ that has n multiply-accumulate (MACC) and corresponding to roughly 2n operations, where $w_i$ and $x_i$ are weight and input of each layer, respectively. 
The gradient descent procedure can be described as repeat {$w_i := w_i + \alpha {dL \over d{w_i}}$} until convergence, so the operation needed is equivalent to one MACC for one parameter in one BP.
We break down the original and morphed models into several components (layers) and analytically compute the operation count needed of each layer in the FP, as listed in Table~\ref{tab:op-fp}.
The detailed descriptions of each layer are in~\cite{nas-resnet,batch-norm}.

\begin{table}[htbp]
	\centering
	\caption{The analytical operation counts of each layer (per image) in the FP. For convolutional layer, the input image dimension is $H_i \times W_i \times C_i$, the output dimension is $H_o \times W_o \times C_o$ and the kernel (filter) size is $K \times K$. For dense layer, the input is $C_i$ and output is $C_o$. Following the convention in~\cite{op-weight}, the operation weight of MACC is 2, the weight of add/subtract/multiply/comparison is 1, the weight of divide/sqrt is 4 and the weight of special operation like exponential is 8. The operation is only an approximation.}
	\begin{tabular}{c|c}
		\hline
		\textbf{Layer} & \textbf{Operation in the FP} \\
		\hline
		Convolutional layer & $MACC = K \times K \times C_i \times H_o \times W_o \times C_o$ \\	
		Dense layer & $MACC = C_i \times C_o$ \\
		Batch normalization & $MACC = Add = Div = H_i \times W_i \times C_i$ \\
		ReLU & $Comparison = H_o \times W_o \times C_o$ \\
		Add layer & $Add = H_o \times W_o \times C_o$ \\
		Max-pooling layer & $Comparison = K \times K \times H_o \times W_o \times C_o$ \\
		Global-pooling layer & $Add = H_i \times W_i \times C_i; Div = C_i$ \\
		Softmax layer & $Exp = Add = Div = C_o$ \\
		\hline
	\end{tabular}
	\label{tab:op-fp}
\end{table}

The analytical analysis of operation computing is more complicated in the BP process.
The convolution in FP can be described as $O_{ij} = \sum\limits_{m=1}^{k-1} \sum\limits_{n=1}^{k-1} X(i-m,j-n) F(m,n)$, where $O_{ij}$ is the output, $X(i-m,j-n)$ is the input and $F(m,n)$ is the filter (kernel).
The partial derivatives of local parameters ($\frac{\partial O}{\partial F}$) and local intermediates ($\frac{\partial O}{\partial X}$) can be easily derived and are used in gradient calculation.
Applying the chain rule, we have the parameters' gradients and the intermediates' gradients by multiplying the loss gradients with the local gradients as Equation~\ref{eq:chain-rule-conv}.
By substituting the derivatives ($\frac{\partial O}{\partial F}$ and $\frac{\partial O}{\partial X}$), we can express the backpropagation as Equation~\ref{eq:backprop-conv}.
Therefore, the total operation needed to calculate all gradients is roughly twice as that in FP.
The total parameter is convolutional layer (without bias) is $K \times K \times C_i \times C_o$, where $K$, $C_i$ and $C_o$ are kernel size, number of input channel and output channel of convolutional layer, respectively.
So the operation needed to update all parameters with gradient descent method is $2 \times K \times K \times C_i \times C_o$.
Consider all steps we can have the total operation in BP shown in Table~\ref{tab:op-bp}.
Since the $K$, $C_i$ and $C_o$ are typically small values in convolutional layers, the total operations in BP in roughly twice that of FP. 

\begin{align} \label{eq:chain-rule-conv}
	\begin{cases}
		\frac{\partial L}{\partial F_{i}} = \sum\limits_{k=1}^{m} \frac{\partial L}{\partial O_{k}} \times \frac{\partial O_{k}}{\partial F_{i}} \\
		\frac{\partial L}{\partial X_{i}} = \sum\limits_{k=1}^{m} \frac{\partial L}{\partial O_{k}} \times \frac{\partial O_{k}}{\partial X_{i}}
	\end{cases}
\end{align}

\begin{align} \label{eq:backprop-conv}
	\begin{cases}
		\frac{\partial L}{\partial F} = Convolution~(Input~X, Loss~gradient \frac{\partial L}{\partial O}) \\
		\frac{\partial L}{\partial X} = Full~Convolution~(Flipped~F, Loss~gradient \frac{\partial L}{\partial O})
	\end{cases}
\end{align}

For dense layer $Y = W^T X + B$, where $Y$, $X$, $W$ and $B$ are output, input, weight and bias of dense layer, respectively.
The intermediates' gradients can be obtained by multiplying the loss gradients ($\frac{\partial L}{\partial Y}$) with the Jacobian matrices of intermediates ($\frac{\partial Y}{\partial X}$).
Similarly, the weights' gradients is $\frac{\partial L}{\partial W} = \frac{\partial L}{\partial Y} \times \frac{\partial Y}{\partial W}$.
In both cases, the operation needed is the same as that in FP.
The bias gradient is $\frac{\partial L}{\partial B} = \frac{\partial L}{\partial Y} \times \frac{\partial Y}{\partial B} = \frac{\partial L}{\partial Y}$ since $\frac{\partial Y}{\partial B} = 1$, therefore resulting no extra operation.
The total parameter in a dense layer (with bias) is $(C_i + 1) \times C_O$ and the total operation needed in the BP of the dense layer is shown in Table~\ref{tab:op-bp}.
Unlike the convolutional layer, the operation of the dense layer in BP is more than tripled of that in FP.
The operation in BP of the rest layers, including batch normalization, activation function (ReLU), element-wise add layer, max-pooling, global-pooling and softmax layer are all ignorable for practical purposes.
We confirmed our analytical method by comparing the results of ResNet-50 on ImageNet~\cite{config-imagenet} with TensorFlow profiler~\footnote{https://www.tensorflow.org/guide/profiler} (only computing operation in the FP) and NVIDIA profiling tools (compute operation in both FP and BP).
In our analytical method, we do not consider any hardware or software optimization that would result in such an effect. 
The operation count from this analytical approach is only related to the neural architecture, hyperparameters configuration and data (like image resolution).
The optimizations that result in less operation will speed up the training or validation processes therefore higher final OPS.
The details of other verification of our OPS measure approach are elaborated in Appendix~\ref{app:flops}.

\begin{table}[htbp]
	\centering
	\caption{The analytical operation counts of each layer (per image) in the BP. The meanings of symbols are the same as in Table~\ref{tab:op-fp}. The total operation needed for calculating the gradients and for updating parameters are summed.}
	\begin{tabular}{c|c}
		\hline
		\textbf{Layer} & \textbf{Operation in the BP} \\
		\hline
		Convolutional layer & $MACC = 2 \times ( K \times K \times C_i \times H_o \times W_o \times C_o )$ \\
		 & $ + ( K \times K \times C_i \times C_o )$ \\
		Dense layer & $MACC = 2 \times C_i \times C_o + ( C_i + 1 ) \times C_o$ \\
		\hline
	\end{tabular}
	\label{tab:op-bp}
\end{table}

\begin{table}[htbp]
	\centering
	\caption{The analytical operation counts of each layer (per image) in FP and BP together. Most operations in ResNet-50 happen in convolutional layer.}
	\begin{tabular}{c|c|c|c|c}
		\hline
		\textbf{Layer} & \textbf{FP} & \textbf{BP} & \textbf{BP/FP} & \textbf{Total} \\
		\hline
		Convolutional & 7.71E09 & 1.52E10 & 1.9755 & 2.29E10 \\
		Dense & 4.10E06 & 1.23E07 & 3.0005 & 1.64E07 \\
		Batch normalization & 7.41E07 & 1.91E03 & 0.00003 & 7.41E07 \\
		ReLU & 9.08E06 & 0 & 0 & 9.08E06 \\
		Max-pooling & 1.81E06 & 0 & 0 & 1.81E06 \\
		Average-pooling & 1.00E05 & 0 & 0 & 1.00E05 \\
		Add & 5.52E06 & 0 & 0 & 5.52E06 \\
		Softmax & 2.10E04 & 0 & 0 & 2.10E04 \\
		Total & 7.81E09 & 1.52E10 & 1.9531 & 2.31E10 \\
		\hline
	\end{tabular}
	\label{tab:op-layer}
\end{table}

Measuring AI-HPC quantitatively is not trivial due to the diversity, sometimes even conflicting, of workloads and metrics.
One single metric like OPS alone may not be sufficient to reflect the AI-HPC computation capabilities considering both the hardware and software.
For example, the data parallelism algorithm that is frequently applied in distributed machine learning will speed up the whole process at a cost of lower average AI accelerator utilization and OPS.
While one can present all relevant metrics separately, we intend to provide a metric to informatively characterize the system's overall performance.
In general, an efficient AI-HPC would perform more computation and result in higher accuracy in less time.
The empirical results~\cite{DL-book} show that the accuracy on the validation dataset increases monotonically and then plateaus over time.
In other words, the error (1 - accuracy) decreases slower and slower over time. 
We would like to compensate for this effect with an increasing changing rate of the metric.
Therefore the absolute value of the partial derivative of the metric with respect to the error should increase with decreasing error.
On the other hand, the partial derivative of the metric with respect to OPS should be independent of OPS to make the computation contribute to the metric uniformly.
We use this metric as a regulated score in our benchmark,  besides the OPS, to quantitatively measure the cross-stack performance of an AI-HPC.
According to the above conditions, we design our regulated score as Equation~\ref{eq:regulated-score}:

\begin{align} \label{eq:regulated-score}
	& Regulated ~ Score = -ln(Error) \times OPS
\end{align}

where $Error \in (0,1)$ and the negative sign keeps the $ln(Error)$ be positive.
Consequently, the regulated score increases faster with lower error and increases linearly with OPS.
For AI systems at the same machine scale but with different software optimizations, the regulated score can reflect the hardware and software co-performance, therefore we also provide it as a complementary result.

\subsection{Fixed and Customizable Configuration}
\label{sec:meth_con}
There are several rules in our benchmark for a fair comparison across various platforms. 
With a "pencil-and-paper" manner~\cite{hpc-benchmark-npb}, our benchmark also has customizable configurations that allow users to optimize the performance.
First, the benchmark should run on a "primary-replica" architecture. 
The primary node is deployed on a strong server without any AI accelerator to dispatch tasks and collect all results from the replica nodes. 
The replica node is composed of one or multiple servers equipped with AI accelerator(s) and can be deployed with or without a container environment.
Both scale-up (multiple AI accelerators on each replica node) and scale-out (one AI accelerator on each replica node) configurations are supported.
Second, the algorithms and search space used for AutoML are fixed, i.e. network morphism for NAS and Bayesian optimization for HPO, with aforementioned operations and hyperparameters. 
The HPO only starts at the fourth round of training on each replica node since the earlier rounds are trained insufficiently, which is also referred to as the warm-up process in this paper.
A predicted accuracy, instead of the actual one, is used in the warm-up process. 
There is also a maximum limit on epoch and patience, which is the number of epochs to wait before early stop if no progress on the validation dataset.
Third, the dataset is fixed to be ImageNet which has 1281167 and 50000 224*224 RGB images for training and validation, respectively. 
We keep the back-end DL framework and most hyperparameters open to further optimization.
This would partially relieve the performance dependency on manual designs and be more independent of the software part of the system.
The data can be formatted in different ways corresponding to the DL framework.
For example, the data loading with TFRecord is more efficient for TensorFlow.
Forth, our benchmark requires the minimum precision to be FP-16 and the maximum error to be 30\%.
A cumulative value of OPS is calculated at each timestamp (0.1-hour step) and the final value is considered as the score.
The summarized configurations are shown in Table~\ref{tab:fix-cus}.

\begin{table}[htbp]
	\centering
	\caption{Fixed and customizable configurations. The customizable setups are predetermined either empirically or experimentally with default values.}
	\begin{tabular}{c|c}
		\hline
		\textbf{Configuration} & \textbf{Fixed and customizable setups} \\
		\hline
		Server arrangement & Fixed: primary-replica \\
		NAS method & Fixed: network morphism \\
		HPO method & Fixed: Bayesian optimization \\
		Dataset & Fixed: ImageNet \\
		DL Framework & Default: TensorFlow \\
		Initial architecture & Fixed: ResNet-50 \\
		Initial weight & Default: method in \cite{config-resnet-weights} \\
		Batch size & Default: 448 \\
		Optimizer & Default: gradient descent with momentum \\
		Learning rate &  Default: 0.1 with linear decay\\
		Loss function & Default: categorical cross entropy \\
		Maximum epoch & Default: 60 \\
		Parallelism & Default: synchronous all-reduce \\
		Parallel data transformation & Default: 48 \\
		Minimum precision & Fixed: FP-16 \\
		Maximum error & Fixed: 30\% \\
		\hline
	\end{tabular}
	\label{tab:fix-cus}
\end{table}

\section{Evaluation}
\label{sec:eva}

\subsection{Setup}
\label{sec:eva_set}
In our preliminary test, We verified our benchmark design (regarding algorithm and implementation) on our local machine with 4 NVIDIA 1080Ti based on CIFAR10 dataset.
For the formal evaluation presented here, we perform it on two large clusters: GPU (NVIDIA V100) cluster and NPU (Huawei Ascend910) cluster.
Both of them are consist of multiple servers each with 2 CPUs and 8 AI accelerators (see Table~\ref{tab:eval-spec} for hardware specifications).
As a modern practice in AI research, we perform the evaluation in containers with the allocated resources and pre-assigned services for the consistency of testing environment.
we utilize Kubernetes to deploy the docker containers that wrap in all the dependencies including the operating system, libraries and workload codes, etc. to provide the running environment.
We use each physical server with the same hardware specifications as either a primary or a replica node for simplicity.
The detailed information of the evaluation environment is shown in Table~\ref{tab:eval-config}.

\begin{table}[htbp]
	\centering
	\caption{Hardware specifications of the two evaluated systems.}
	\begin{tabular}{c|c|c}
		\hline
		\textbf{Components} & \textbf{GPU cluster} & \textbf{NPU cluster}\\
		\hline
		Processor & Intel skylake 6151 & Huawei Kunpeng 920 \\
		Memory & 2667MHz DDR4 512 GB & 2933MT/s DDR4 2048 GB \\
		AI Accelerator & \makecell[c]{NVIDIA NVLink V100} & \makecell[c]{Huawei Ascend 910} \\
		Storage & NVMe 5 TB & NVMe 5 TB\\
		Ethernet network & InfiniBand 100 Gb/s & InfiniBand 100Gb/s\\
		\hline
	\end{tabular}
	\label{tab:eval-spec}
\end{table}

\begin{table}[htbp]
	\centering
	\caption{The evaluation environments of the two evaluated systems.}
	\begin{tabular}{c|c|c}
		\hline
		\textbf{Components} & \textbf{GPU cluster} & \textbf{NPU cluster} \\
		\hline
		Allocated resources & \makecell[c]{30 CPU cores \\ 128 GB memory \\ 8 NVIDIA V100} & \makecell[c]{191 CPU cores \\ 512 GB memory \\ 8 Huawei Ascend910}  \\
		\hline
		Environment & \makecell[c]{Ubuntu 16.04 \\ docker 18.09 \\ SLURM 15.08 \\ TensorFlow V2.2 \\ CUDA V10.1 \\ Python 3.5} & \makecell[c]{Ubuntu 18.04 \\ docker 19.03 \\ SLURM 17.11 \\ MindSpore V1.0 \\ CANN V20.1 \\ Python 3.7} \\
		\hline
	\end{tabular}
	\label{tab:eval-config}
\end{table}

\subsection{Performance}
\label{sec:eva_perf}
We run the benchmark on various scales of machines from 10 replica nodes with 80 GPUs up to 512 replica nodes with 4096 NPUs.
All the intermediate results including the generated architectures, hyperparameter configurations, accuracy at each epoch and timestamps are recorded in log files.
Once the benchmarking process is finished, we run the data analysis toolkit to calculate the score along with other complementary results utilizing all the recorded information and then create a report.

\begin{figure}[htbp]
	\centering
	\includegraphics[width=0.5\textwidth]{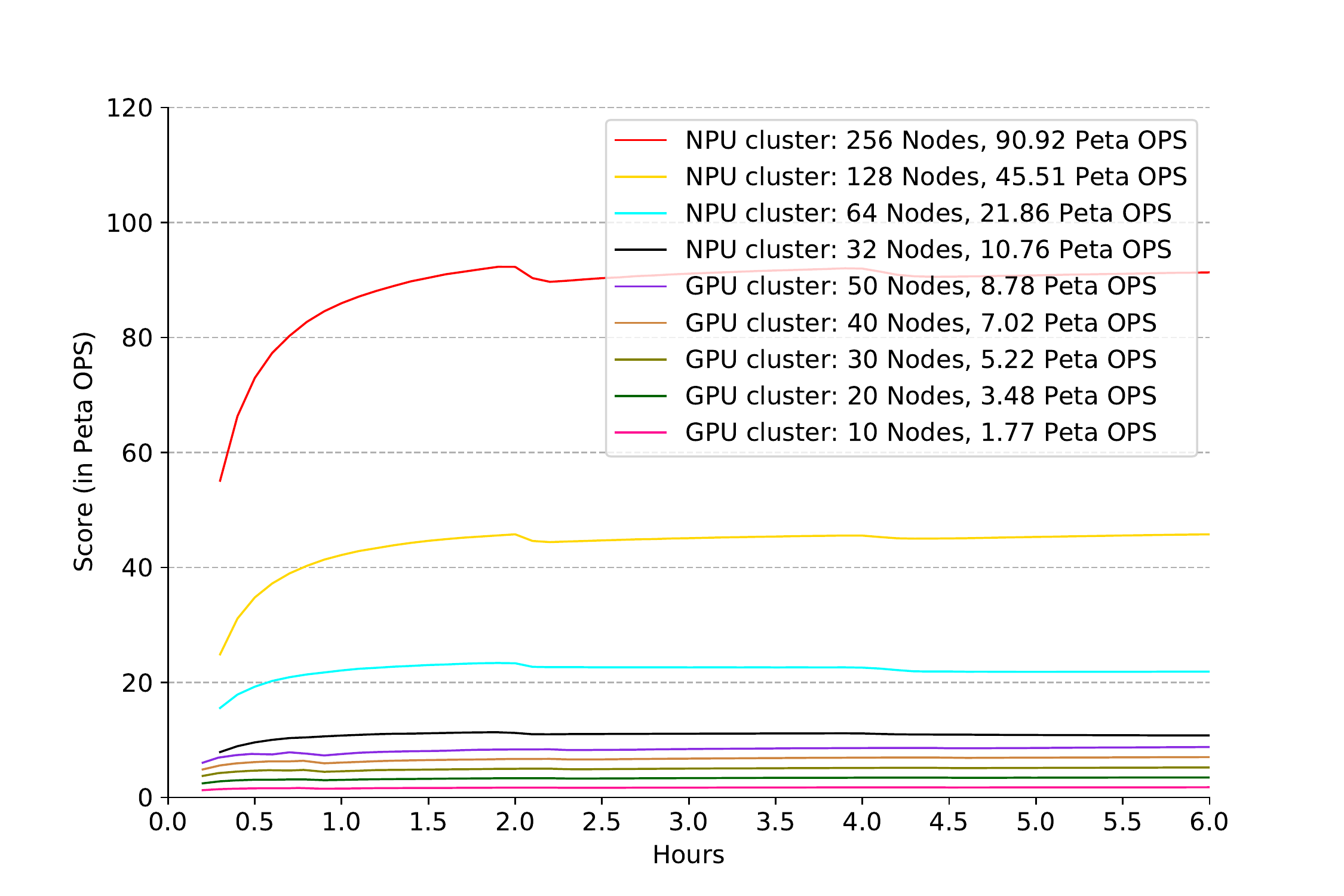}
	\includegraphics[width=0.5\textwidth]{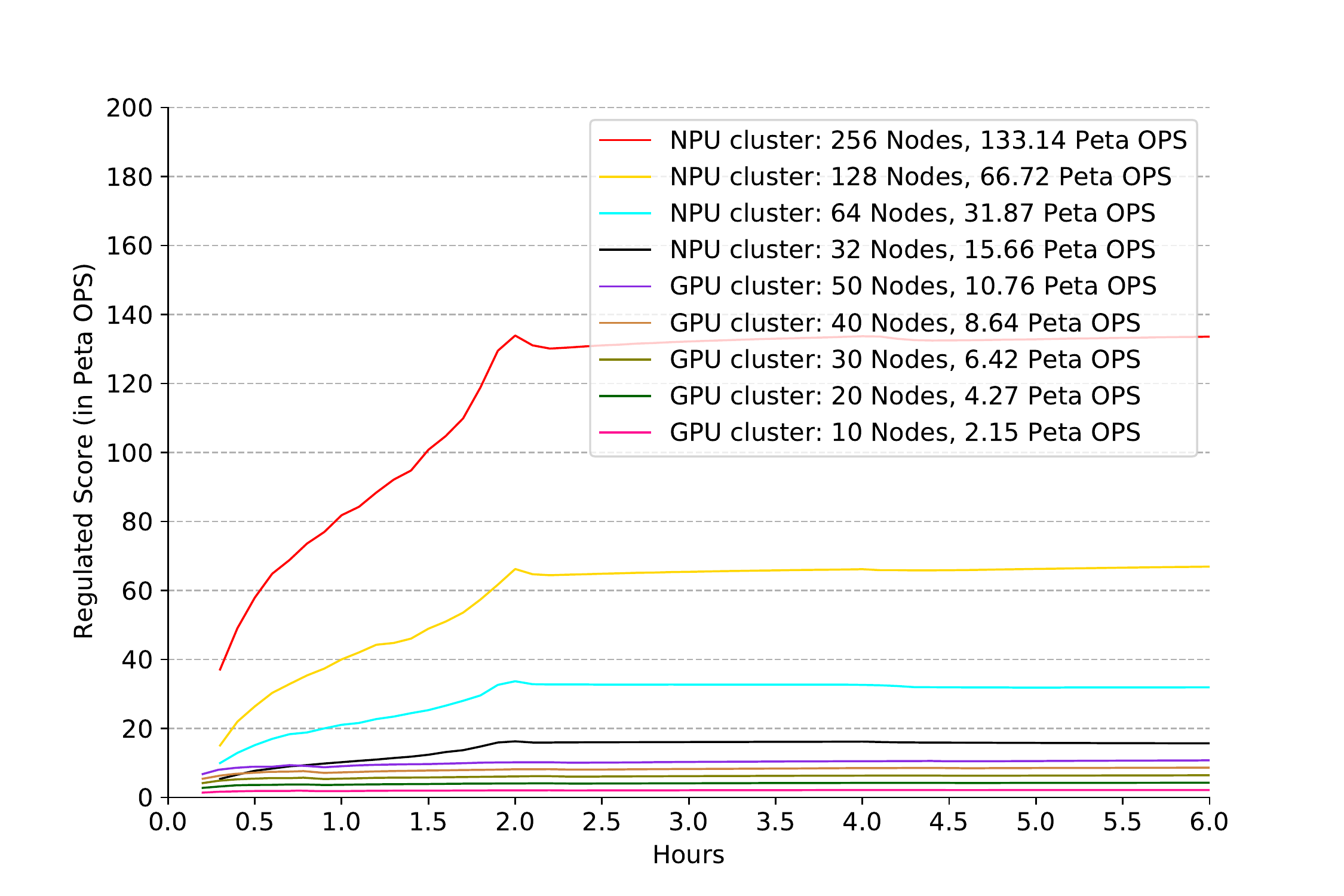}
	\caption{The benchmark scores and the regulated scores (both in Peta OPS) over time of evaluations with different scales of cluster nodes. The cumulative OPS is converged and increases steadily after the initial warm-up phase and we report the final value as the benchmark score, shown in the labels. The results show robustness and stability of our benchmark. The regulated scores have similar behaviors as the scores.}
	\label{fig:result-score}
\end{figure}

In this paper, we limit our evaluation in two major characteristics of the benchmark: stability and scalability. 
As for stability characteristic, within the pre-assigned hours on various types and scales of AI accelerators, the cumulative OPS is calculated and shown in Fig.~\ref{fig:result-score} as score.
As we can see that in both two clusters, the cumulative OPS converges and increases steadily.
The regulated score in Fig.~\ref{fig:result-score} also converges since it is essentially just OPS multiplied with the model performance as a coefficient.
The regulated score has similar behavior as score.

\begin{figure}[htbp]
	\centering
	\includegraphics[width=0.5\textwidth]{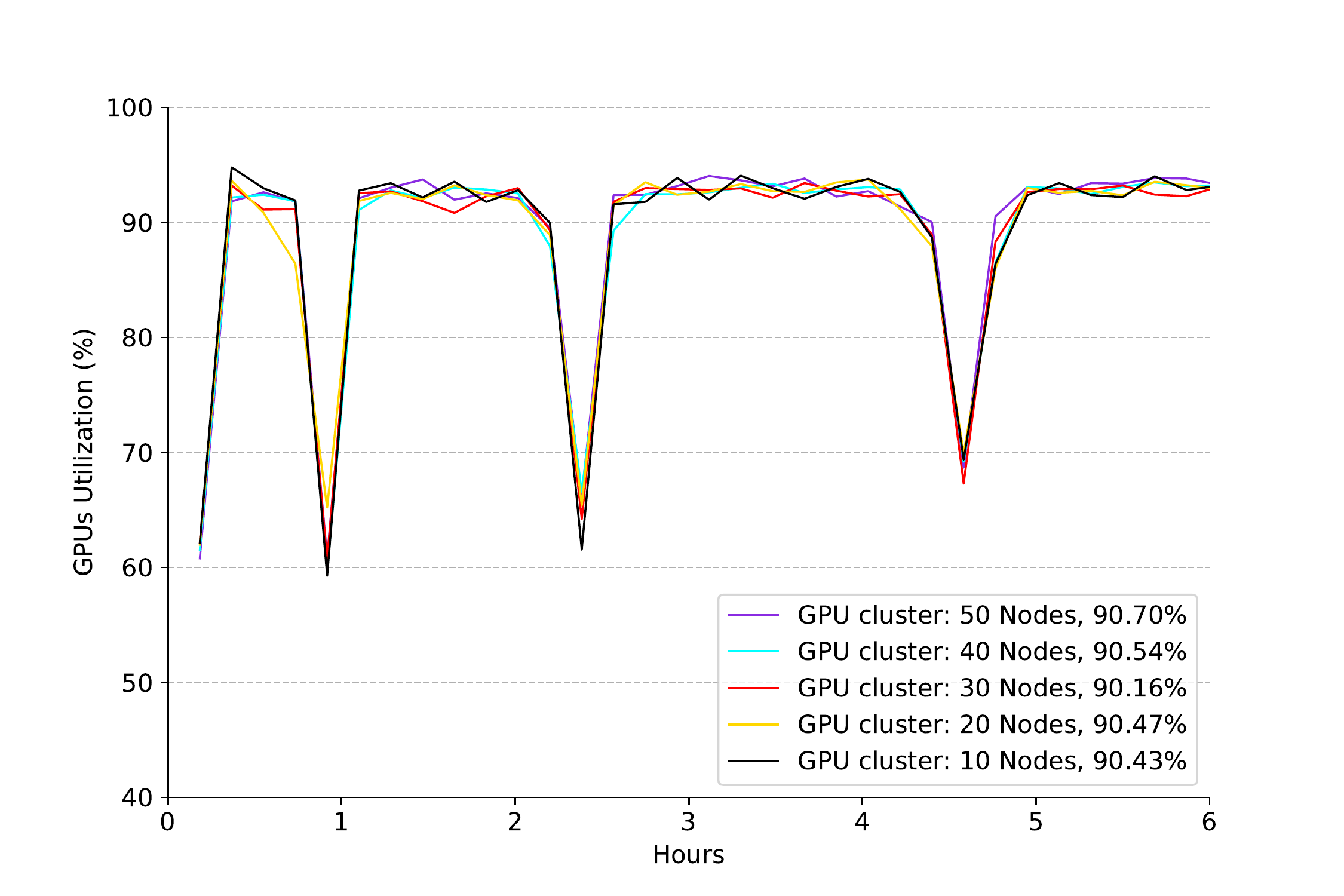}
	\includegraphics[width=0.5\textwidth]{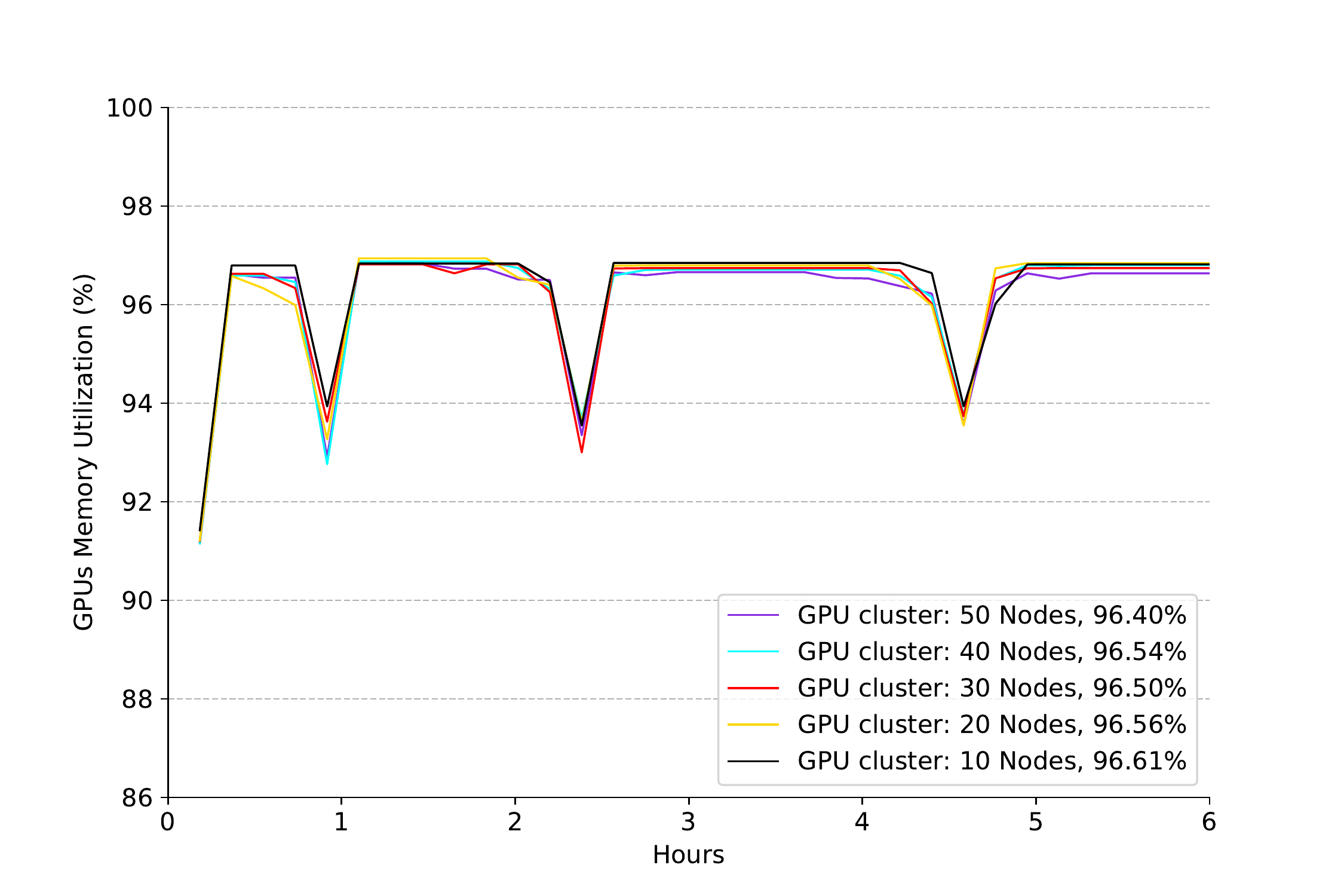}
	\caption{The GPUs utilization and their memory utilization of evaluations with different scales of machines measured by NVIDIA profiling tool. The average values are shown in the labels. The utilization drops during the inter-phase between the training stages come from the data loading and computational graph compilation etc.}
	\label{fig:gpu-util}
\end{figure}

To ensure the stability, we also monitor the GPU performance during the benchmarking process.
We use NVIDIA System Management Interface (nvidia-smi~\footnote{https://developer.nvidia.com/nvidia-system-management-interface}) to track the GPU utilization to show the percentage of time during which one or more kernels are occupied, along with the GPU memory utilization during the same time period.
We developed a toolkit to extract real-time information with 30 seconds sampling interval during the entire running time.
As shown in Fig.~\ref{fig:gpu-util}, the GPU utilization and memory occupancy are both high during the training phase with the default benchmark configuration (for NVIDIA V100).

\begin{figure}[htbp]
	\centering
	\includegraphics[width=0.5\textwidth]{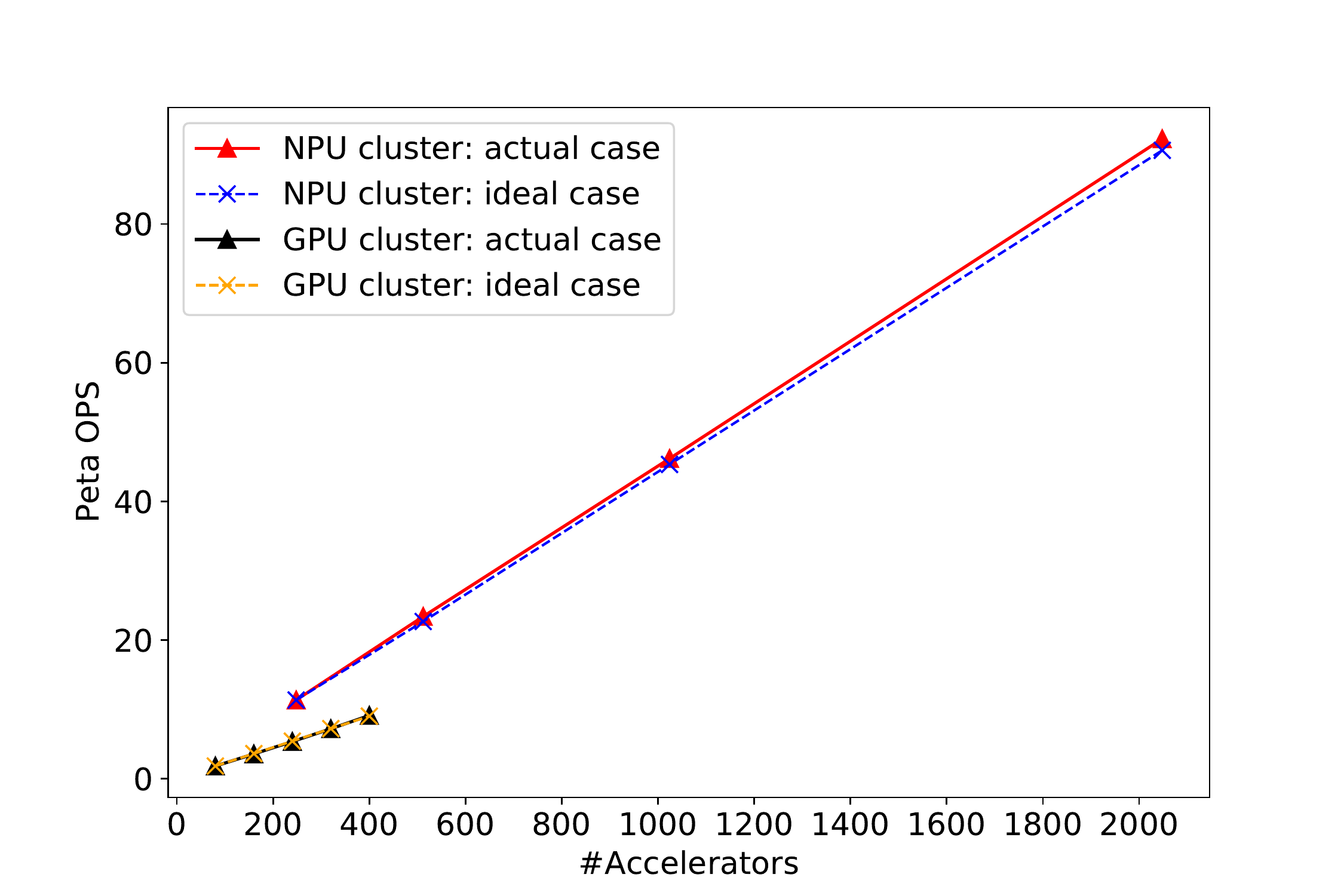}
	\caption{The computing speeds (in Peta OPS) over different number of AI accelerators. The benchmark shows near-linear weak scalablity on the two evaluated systems. Due to the optimization of the benchmark configurations and system fluctuation, super-linear effect appears occasionally.}
	\label{fig:weak-scaling}
\end{figure}

As described in Section~\ref{sec:meth}, the job size increases when the number of processing units increases.
Hence, as for scalability characteristic, the weak scaling test is performed and the result is shown in Fig.~\ref{fig:weak-scaling}.
The benchmark shows near-linear weak scalablity on the two evaluated systems, which implies that our benchmark is able to evaluate even bigger systems such as future exascale supercomputers.
Due to the optimization of the benchmark configurations and system fluctuation, super-linear effect appears occasionally.

\section{Conclusion}
\label{sec:con}
The rise of the convergence of AI and HPC reveals new challenges in benchmarking the state-of-the-art and future large scale clusters for AI purposes.  
We review the current HPC and AI benchmarks and explain why they do not address all the challenges.
We choose AutoML, a highly scalable and representative AI application, as our benchmark workload and implement the algorithms in a highly parallel manner.
We also propose an analytical approach that is independent of DL frameworks and other software implementations to estimate the computation operation rate during training and validation processes.
We utilize this rate as the benchmark score to construct the benchmark score to quantitatively measure the machine computing power on AI applications.
We evaluate the benchmark on different types and scales of systems with a large dataset and verify the benchmark's stability and scalability.
Moreover, the simple metric design allows us to compare and rank machines from small clusters to large AI-HPC easily.

\section*{Acknowledgment}

We thank xxxxxx

\section{Appendix: OPS Calculation}
\label{app:flops}
We compare our analytical approach of operations computing with the TensorFlow profiler (tf.profiler~\footnote{\url{https://www.tensorflow.org/api_docs/python/tf/compat/v1/profiler/ProfileOptionBuilder}}) and NVIDIA profiling tool (nvprof).
The tf.profiler can only count operations in the FP.
The nvprof can trace the GPU activity and use the kernel replay to ensure all requested profile data including operation counts of adds, multiplies, multiply-accumulates, and special operations.
The profiling process with nvprof is prohibitively expensive therefore we need an approach to speed-up the process.

\begin{table}[H]
	\caption{The comparison of operation counts of each layer of ResNet-50 on ImageNet (per epoch, batch size=1, input shape=224*224*3) with different approaches. The difference of FPs in training and validation stage comes from the data size.}
	\label{tab:op-compare}
	\centering
	\begin{tabular}{l|l|l|l}
		\hline\noalign{\smallskip}
		\textbf{Procedure} & \textbf{tf.profiler} & \textbf{nvprof} & \textbf{analytical} \\
		\noalign{\smallskip}\hline\noalign{\smallskip}
		FP (training) & 9.97E15 & 1.02E16 & 1.00E16 \\
		BP (training) & - & 2.10E16 & 1.95E16 \\
		BP / FP (training) & - & 2.0603 & 1.9533 \\
		Total (training) & - & 3.12E16 & 2.95E16 \\
		FP (Validation) & 3.89E14 & 3.98E14 & 3.90E14 \\
		Total (training + validation) & - & 3.16E16 & 2.99E16 \\
		\noalign{\smallskip}\hline
	\end{tabular}
\end{table}

Fortunately, we can utilize the iterative nature of DL computation and sample the profiling process based on a small partition of data.
This is only an approximation since the operations varies with the hyperparameter configurations.
Table~\ref{tab:op-compare} shows the operations of ResNet-50 layers on ImageNet with the 3 approaches.
The operations in BP are consistent between our analytical approach and nvprof, and the operations in FP are consistent among all three approaches.

\bibliography{references} 

\end{document}